\documentclass[a4paper, 12pt]{report}			
\usepackage[english]{babel}
\usepackage[utf8]{inputenc}	

\usepackage[usenames,dvipsnames,svgnames,table,x11names]{xcolor}
\usepackage{graphicx}
\usepackage{tikz}
\usepackage[plainpages=false,pdfpagelabels,hidelinks,breaklinks]{hyperref}
\usepackage{hyphenat}
\usepackage[figure]{hypcap} 
\usepackage[T1]{fontenc}
\usepackage[disable]{todonotes}
\usepackage{geometry}

\usepackage{verbatim}
\usepackage[normalem]{ulem}
\usepackage{soul}
\usepackage{soulutf8}
\sethlcolor{yellow}
\usepackage{amsmath}
\usepackage{multicol}
\usepackage{multirow}
\definecolor{light-gray}{gray}{0.9}
\usepackage{pdfpages}
\usepackage{textcomp}
\usepackage{chngcntr}
\counterwithout{figure}{chapter}
\counterwithout{table}{chapter}

\usepackage{listings}
\renewcommand\lstlistlistingname{Quellcodeverzeichnis}
\definecolor{lstbackgroundcolor}{rgb}{0.97, 0.97, 0.97}
\definecolor{lstkeywordcolor}{rgb}{1, 0, 1}
\definecolor{lstHellblau}{rgb}{0,0.4,0.7}
\lstdefinelanguage{myIni}
{
	basicstyle=\normalfont\rmfamily,
	alsoletter={=},
	morekeywords={=},
	keywordstyle=\color{lstkeywordcolor},
	numbers=left,
	numberstyle=\footnotesize,
	captionpos=t,
	showstringspaces=true,
	morecomment=[s][\color{blue}]{[}{]},
	morecomment=[l][\color{ForestGreen}\itshape]{;},
	xleftmargin=.06\textwidth,
	xrightmargin=.06\textwidth, 
	backgroundcolor=\color{lstbackgroundcolor},
	frame=single,
	extendedchars=true,
	stepnumber=1,
	aboveskip=\parskip,
	belowskip=\parskip,
  belowcaptionskip=1em,
	boxpos=c,
	breaklines=true,
	linewidth=\textwidth,
}
\lstdefinelanguage{myC++}
{
  language=C++, 
	basicstyle=\normalfont\rmfamily,
	numbers=left,
  morekeywords={TRUE,FALSE},
  emph={ULONG,BOOL,AXIS_REF,USHORT,INT,FLOAT,DOUBLE},
  emphstyle={\color{lstHellblau}},
	numberstyle=\footnotesize,
	captionpos=t,
	showstringspaces=true,
	xleftmargin=.06\textwidth,
	xrightmargin=.06\textwidth, 
	backgroundcolor=\color{lstbackgroundcolor},
  keywordstyle=\color{blue}\bfseries,
  commentstyle=\color{ForestGreen}\itshape,
  stringstyle=\color{red!50!brown},
	frame=single,
	extendedchars=true,
	stepnumber=1,
	aboveskip=\parskip,
	belowskip=\parskip,
  belowcaptionskip=1em,
	boxpos=c,
	breaklines=true,
	linewidth=\textwidth,
  tabsize=2,
}
\lstdefinelanguage{myGolang}
{
	morekeywords=[1]{define, define-syntax, define-macro, lambda, define-stream, stream-lambda},
	morekeywords=[2]{begin, call-with-current-continuation, call/cc,
		call-with-inpushowstringspacest-file, call-with-output-file, case, cond,
		do, else, for-each, if,
		let*, let, let-syntax, letrec, letrec-syntax,
		let-values, let*-values,
		and, or, not, delay, force,
		quasiquote, quote, unquote, unquote-splicing,
		map, fold, syntax, syntax-rules, eval, environment, query },
	morekeywords=[3]{import, export},
	alsodigit=!\$\%&*+-./:<=>?@^_~,
	sensitive=true,
	morecomment=[l]{;},
	morecomment=[s]{\#|}{|\#},
	morestring=[b]",
	basicstyle=\normalfont\rmfamily,
	keywordstyle=\bf\ttfamily\color[rgb]{0,.3,.7},
	commentstyle=\color[rgb]{0.133,0.545,0.133},
	stringstyle={\color[rgb]{0.75,0.49,0.07}},
	upquote=true,
	breaklines=true,
	breakatwhitespace=true,
	backgroundcolor=\color{lstbackgroundcolor},
	literate=*{`}{{`}}{1},
	numbers=left,
	numberstyle=\footnotesize,
	captionpos=t,
	showstringspaces=true,
	xleftmargin=.06\textwidth,
	xrightmargin=.06\textwidth, 
	frame=single,
	extendedchars=true,
	stepnumber=1,
	aboveskip=\parskip,
	belowskip=\parskip,
	belowcaptionskip=1em,
	boxpos=c,
	linewidth=\textwidth,
	tabsize=2,
}
\lstdefinelanguage{myConsole}
{
	basicstyle=\normalfont\rmfamily,
	alsoletter={=},
	morekeywords={=},
	keywordstyle=\color{lstkeywordcolor},
	numbers=left,
	numberstyle=\footnotesize,
	captionpos=t,
	showstringspaces=true,
	morecomment=[l][\color{ForestGreen}\itshape]{;},
	xleftmargin=.06\textwidth,
	xrightmargin=.06\textwidth, 
	backgroundcolor=\color{lstbackgroundcolor},
	frame=single,
	extendedchars=true,
	stepnumber=1,
	aboveskip=\parskip,
	belowskip=\parskip,
	belowcaptionskip=1em,
	boxpos=c,
	breaklines=true,
	linewidth=\textwidth,
	backgroundcolor=\color{lstbackgroundcolor},
}
\lstdefinelanguage{myPython}{
	keywords={typeof, null, catch, switch, in, int, str, float, self},
	keywordstyle=\color{ForestGreen}\bfseries,
	ndkeywords={boolean, throw, import},
	ndkeywords={return, class, if ,elif, endif, while, do, else, True, False , catch, def},
	ndkeywordstyle=\color{BrickRed}\bfseries,
	identifierstyle=\color{black},
	sensitive=false,
	comment=[l]{\#},
	captionpos=b,
	morecomment=[s]{/*}{*/},
	commentstyle=\color{purple}\ttfamily,
	stringstyle=\color{red}\ttfamily,
	breaklines=true,	
}

\usepackage{titlesec}

\titleformat{\paragraph}
{\normalfont\normalsize\bfseries}{\theparagraph}{1em}{}
\titlespacing*{\paragraph}
{0pt}{3.25ex plus 1ex minus .2ex}{1.5ex plus .2ex}

\usepackage[onehalfspacing]{setspace}	
\usepackage{epstopdf}
\usepackage[backend=biber,bibencoding=utf8,style=alphabetic,sorting=none]{biblatex}		
\usepackage[babel,german=guillemets]{csquotes}
\newcommand{\ml}{machine learning}
\newcommand{\hp}{hyperparamter}

\begin{document}
\includepdf[pages={1}]{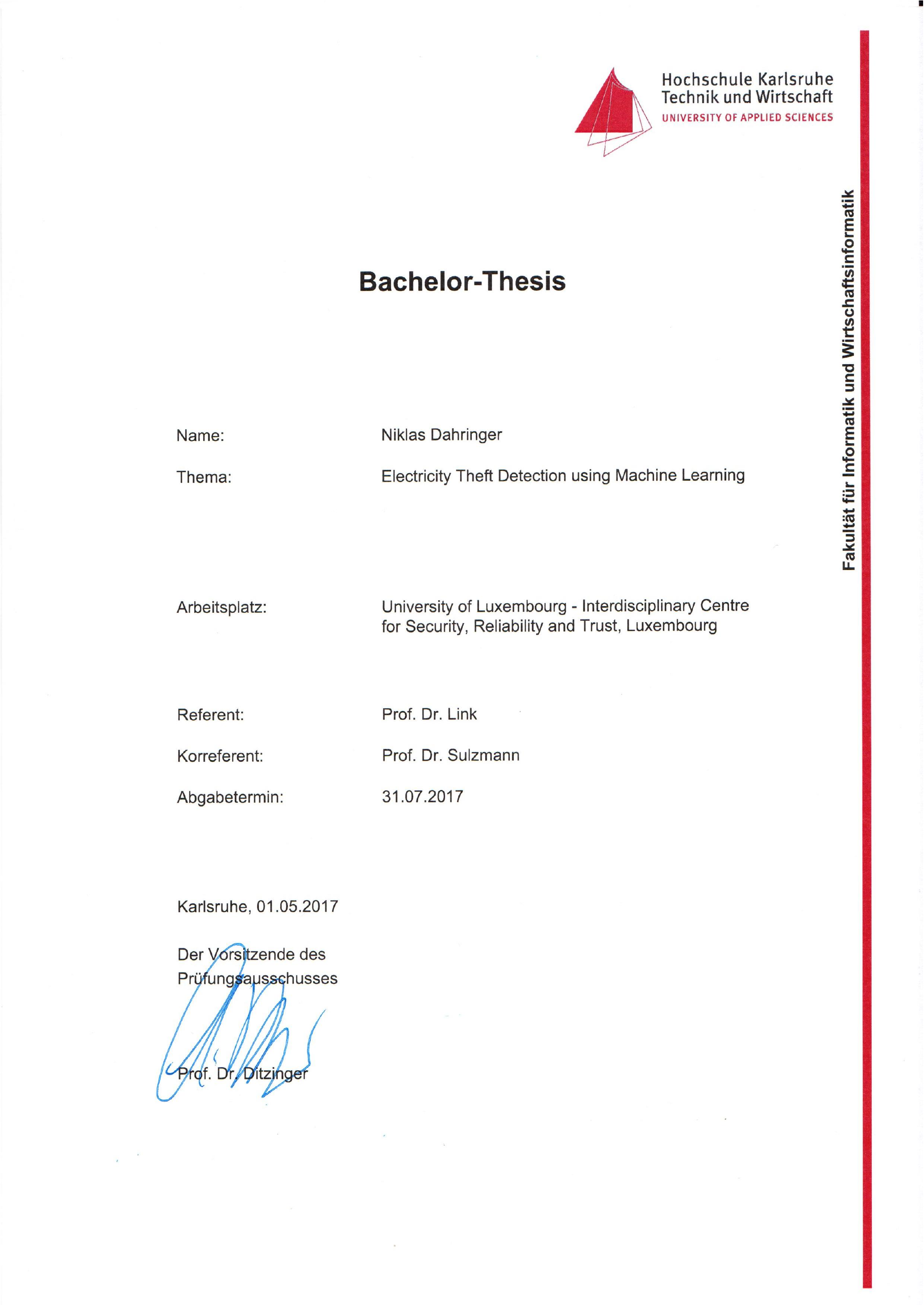}
\chapter*{Declaration}
\thispagestyle{empty}
I herewith certify that all material in this report which is not my own work has been properly
acknowledged.\\
\newline
\newline
Niklas Dahringer

\clearpage
\begin{abstract}
Non-technical losses (NTL) in electric power grids arise through electricity theft, broken electric meters or billing errors. They can harm the power supplier as well as the whole economy of a country through losses of up to 40\% of the total power distribution. For NTL detection,  researchers use artificial intelligence to analyse data. This work is about improving the extraction of more meaningful features from a data set. With these features, the prediction quality will increase.
\end{abstract}

\chapter*{Acknowledgements}
\thispagestyle{empty}

I would like to thank my home university, the Karlsruhe University of Applied Sciences, and my host university, the University of Luxembourg, to make this work possible. Also, it is my pleasure to thank the CHOICE Technologies Holding S\`{a}rl for providing the interesting subject and the data. Furthermore, I very much appreciate the support of my supervisor Professor Dr. Norbert Link. Additionally, I want to thank Patrick Oliver Glauner for his professional support.

\clearpage
\newpage
\pagenumbering{Roman}
\setcounter{page}{2}
\tableofcontents



\newpage
\pagenumbering{arabic}
\newpage
\chapter{Introduction}
This work is created in cooperation with the Karlsruhe University of Applied Sciences  (HsKa) and with the Interdisciplinary Centre for Security, Reliability and Trust, University of Luxembourg.\\
The focus of this work is to improve the quality of the detection of non-technical losses in electric power grids.

\section{Non-technical losses}
The loss of electricity  is still a problem for electricity suppliers. Losses have different reasons: There are technical reasons, like the internal resistance of power grid components, as happens in transformers, generators and transmission lines.  However, these are not the only  drawbacks for the power suppliers, as there is also non-technical loss (NTL).
NTLs could be electricity meter manipulation, bypassing meters, or bribing meter readers to reflect a lower power consumption.
Also, possible cases of NTL are broken or faulty meters, unmetered supply, technical and human errors in meter readings, data processing, and billing.\\
Mainly, NTLs cause a drop in revenue for the power suppliers.
Also, uncontrolled power deprivation leads to a decline of the stability of electric power grids.
This instability can cause disadvantages for the whole economy of a country. In some countries like Brazil, India, Malaysia or Lebanon, NTLs account for 40\% of the total power distribution. Even in developed countries NTL is a problem.\\
Two approaches exist to detect NTL: The first way is to determine the total power consumption for the whole power grid including its components. Problems appear through changes or faults in the network. Also, it would  be necessary to record the status of all elements in the power grid. \\
A much better approach, which this work employs, is to use machine learning to detect irregularities in consumption data and, based on the results, to decide to perform an inspection or not.~\cite{survey-glauner}

\section{Machine learning}
One step among many others in this work is the use of machine learning. Machine learning is when a computer acts without being directly programmed for this task, according to Andrew Ng~\cite{ng-mlexplain}. \\
In other words, {\ml} is used to make decisions. Machine learning is used to find a solution with a lack of the concrete implementation, which will be compensated with a larger data set. The operator does not directly implement the final decision process of the program; it is learned from the data. More specifically, the idea is to extract characteristics from the data and to train the program parameters with these features.\\
With this trained program, it is possible to make predictions about similar data, albeit not comprehensively. Machine learning will find some patterns in the training data and, using these models to make approximations for future data possible.\\
An intelligent program should be able to react to changes very easily. It would be a huge disadvantage to reimplement the whole program because of small shifts in the data. With {\ml} and its potential to retrain  to obtain better precision, it becomes needless to provide a concrete solution process to train the program. Because of this behaviour, to learn even from changing data, {\ml} belongs to  artificial intelligence.\\

Besides the {\ml} model parameters, the so-called hyperparameters are another set of parameters that are given to describe complex properties of the model before the actual training begins, for example, the maximum depth of the trees of a random forest or the limit of the maximum number of possible leaf nodes. The {\ml} model parameters adjust during the training to fit the patterns found in the training data. However, it is also possible to choose the best-fitting hyperparameters automatically with a random search function plus a cross-validation; for a detailed description see Section~\ref{sec:methods}. \\

It is important to be aware that these parameters, learned with the help of patterns, will and should never entirely fit the training data, because after the training of the parameters, test data sets have different characteristics; this also includes outliers. If the parameters would precisely fit the training data, it could cause problems for the later use of the model. The decisionmaking with the test set would be a failure because the parameters would only fit the training data.  This issue is known as overfitting; to avoid this, it is necessary that the parameters do not fit the training set too closely.\\
Another problem would be underfitting, meaning that too many characteristics from the training data are ignored, leading to a too simplified model.\\

Machine learning can be used for diverse types of data like data base entries, pictures, audio files, etc.~\cite[41-44]{intro-ml}\cite{underfitting-web}.

\subsection{Classification and regression}
After the training, the {\ml} model can perform decisions like classification. This is about recognizing patterns in the data and matching the data to a distinct class. Classes are the number of possible results, e.g., the number of different exisiting categories. \\ 
Another use case, as opposed to the classification used in this work, is the regression, which is used to describe a relation between the classes rather than matching them to distinct class. Therefore, float numbers are often used to express the regression.
Both these functions, regression and classification, are part of the supervised learning~\cite[45-49]{intro-ml}.\\

\subsection{Classifiers}
\subsubsection{Random forest}
The random forest algorithm belongs to the category of ensemble learning methods of machine learning algorithms and consists of other weaker ones.
The random forest classifier is known for its fast execution.
An interesting part is the training of these algorithms: every tree of the forest is trained separately with a random subset of the training data. So, every single tree can predict a part of the data well, and this helps to avoid overfitting. 
It is common to use randomization to receive a good distribution of the subsets from the training data. Randomization is also used for the hyperparameter calculation, see Section~\ref{sec:methods}.\\
The combination of the classification results from every tree leads to a result that is less affected by biases.~\cite{decisiontree-ml, randomforest-ho}

\subsubsection{Decision tree}
A decision tree is used for classification as well as for regression.
Regarding classification, decision trees are famous for their speed and accuracy. Well-known implementations of decision trees are ID3 and C4.5.
During the training a decision tree grows until time it makes a decision. While doing so, its important to adjusts the training set to avoid overfitting~\cite{decisiontree-ml, decisiontree-quinlan}.

\subsubsection{Support vector machine}
A support vector machine (SVM) is a supervised machine learning model that is used for classification and as well for regression. The idea is to separate classes with the help of hyperplanes, which are determined by the support vectors~\cite{svm-intro}.
An SVM is able to map data into a higher dimension to handle complex separations~\cite[138]{vapnik2000nature}. However, the execution with large data sets will slow it down. Therefore, linear implementations of SVMs are often used~\cite{svm-speed}. Furthermore, an SVM is more robust against overfitting than other classifiers like neural networks~\cite{svm-overfitting}.

\subsubsection{Gradient-boosted tree}
The gradient-boosted tree is a supervised machine learning model. Additionally, it is an ensemble algorithm that consists of weaker algorithms like decision trees. A loss function prevents losses during the adding of trees one by one~\cite{chen2016xgboost, gradientboost-ml}.

\section{Goals}
The goal of this project is to get as much quality features as possible out of the consumption time series. Therefore, attempts are made to gain more meaningful features from the data, taking into account the noise in the provided real-world data. Additionally, this work focuses on the improvement of feature extraction and selection.\\
To verify the result, different classifiers are trained with the gained features and eventually compared against each other.
For all of this, the tsfresh library and additional Python modules are used.\\

\chapter{Related work}
The SEDAN research group from the Interdisciplinary Centre for Security, Reliability and Trust (SnT) of the University of Luxembourg, and the CHOICE Technologies Holding S\`{a}rl worked together for a longer time on the topic of NTL detection. They publish diverse papers about this domain.\\

A highly informative paper is the \textit{The Challenge of Non-Technical Loss Detection Using
Artificial Intelligence: A Survey}~\cite{survey-glauner}, which gives an overview about different topics, namely, different types of features for NTL detection, like monthly consumption, smart meter consumption, master data, which could be described as metadata because they insist on the name and address of the customer, the feeder voltage, and perhaps climate data. Another feature type would be the creditworthiness, which ranks the payment morale or the income of a customer. 
Also, the survey covers different detection approaches like expert systems, fuzzy systems, neural networks, support vector machines and others. In addition, the paper assesses other papers about NTL detection and their quality. The size of the data sets and the scoring methods for the NTL detection quality are criticized.\\

A data set of circa 22K customers is used in \cite{costa2013fraud} for training a neural network. It uses the average consumption of the previous 12 months and other customer features such as location, type of customer, voltage and whether there are meter reading notes during that period. On the test set, an accuracy of 0.8717, a precision of 0.6503 and a recall of 0.2947 are reported.\\

Consumption profiles of 5K Brazilian industrial customer profiles are analyzed in \cite{ramos2012identification}. Each customer profile contains 10 features including the demand billed, maximum demand, installed power, etc. In this setting, a SVM slightly outperforms $k$-nearest neighbors (KNN) and a neural network, for which test accuracies of 0.9628, 0.9620 and 0.9448, respectively, are reported. \\

Glauner et al.~\cite{glauner2016large} compare different classifiers on a real-world data set of 100,000 customers. Furthermore, they change the imbalances to check the behaviour of the classifiers. As a feature they used the daily average consumption from the data and therewith reached a prediction quality slightly above random guess. \\
This work shows that it must be possible to improve the prediction quality through enhancing the quality of the used features.\\

Detecting NTL with provider-independent data is the domain of a paper by Meira et al.~\cite{distilling-jorge}. Therein they compute different features with due consideration of the temporal, local and similarity criteria. Subsequently, they train three classifiers and evaluate the results.\\
Of course, using metadata helps to improve the prediction performance, but this work will focus on improving features from the consumption time series.

\chapter{Technologies used}
\section{Python Programming language}
The programming language which is used for this project is Python. Python is an interpreter language which also is a high level language. It has been developed by Guido van Rossum~\cite[12]{pythonbook-intro} and is maintained by the Python Software Foundation~\cite{python-psf}. 

A very special characteristic compared to other programming language is the separation of code blocs. Separations are used for inner function or commands which contains other operation like the if operator or diverse kinds of loops. Very common in other languages is to use curly brackets at begin and at end of a block or using for this separation task a specific key word. As a separation   indentation are used in Python which are readable for humans and for the interpreter. Such indentation is composed of four spaces.~\cite{pythonwiki-indentation}

In the year 2008 the version 3.0 was released which brought explicit changes in the syntax of Python~\cite{pythonwiki-version}. Because it is not  easy to migrate every code to 3.X the version 2.7 got an extended maintenance~\cite{change-version}. This leads to the fact that now both versions are in use.

For Python exists numerous packages which extend the function of the original programming language, for example the in this work used packages NumPy, pandas and Dask.
\subsection{NumPy}
A well-known package in Python is NumPy which is advertised for scientific work.
NumPy is predesignated for mathematical tasks, it  provides  multidimensional arrays, sorting, basic statistical operations and diverse other function for different purposes~\cite{scipy-numpy}. Furthermore, it offers  many functions regarding the multi dimensional arrays. Often, during the evaluation of the results from this work, the  random number generator was used.~\cite{numpy-home, pythonwiki-numpy}

\subsection{pandas}
During the whole preprocessing and in the library tsfresh itself the Python package pandas is used. Pandas is conceived for the manipulation and analyses of time series and data tables. 
It is able to work with named and unnamed data which are acquired from SQL data sets, Excel spreadsheets or any other data types. The pandas package orients oneself by the NumPy package.
\\
Pandas has two main structures \texttt{Series} and \texttt{DataFrame}, the first one  has only one dimension the second one has two dimensions.
This package can handle  special data formats, for example sometime in a \texttt{DataFrame} occurs  unreadable values such values will be represented as \texttt{nan} (NotaNumber), also pandas can deal with the NumPy data types inf (infinity) or -inf.
Very useful is the groupby function which groups  entries by identical values. On this basis, is it possible to perform a variety of different operations on the data. This is  called in pandas split-apply-combine operations, too.
Also  important is the ability to slice \texttt{DataFrame}s and \texttt{Series} which takes effect for  horizontal and vertical slicing, orientated on  label, column number, on  the index position or index number.
For merging and joining different \texttt{DataFrame}s a lot of parameters are provided to perform different types of joins, which also was used to merge \texttt{DataFrame}s in this work.~\cite{pandas-info}

\subsection{Dask}
Dask is a  library for parallel computing it is available as a Python package.
The library  consists out of two components,  one is for the dynamic task scheduling.  The  other one  part contains all the different collections like constructs from pandas and NumPy.
This is the  main advantage the NumPy and pandas like structures and functions. Which makes  it easier to parallelize the source code because it's familiar to  the common NumPy and pandas packages.~\cite{dask-info}\\
Furthermore Dask provides a large set of parameter for the optimization of the parallelization and adaptation for different systems.
It is possible to load external data direct as a Dask \texttt{DataFrame} or to convert a pandas \texttt{DataFrame} to a Dask \texttt{DataFrame}, therefore are some meta data required which describe the original \texttt{DataFrame}.
The  conversion from a pandas to a Dask structure or the conversly process is a overhead in  this library however it is necessary to do this for some  in Dask missing \texttt{DataFrame}-operations.~\cite{dask-create}\\
Also Dask offers the chance to parallelize code which aren't covered by the Dask functions. Therefore exists the delay function which wraps the modified original function.~\cite{dask-delayed}

\subsection{scikit-learn}
scikit-learn~\cite{scikit-learn} is python library for machine learning.  Which is published as open source. The library is qualified for  data mining and data analysis, therefore it contains  different function  for classification, regression, clustering, dimensionality reduction, model selection and preprocessing. To realise these funcionality it uses different python library like NumPy, SciPy and matplotlib.~\cite{sklearn-front}\\
The project launch of scikit-learn was in 2007 during a Google Summer of Code.  Subsequently Matthieu Brucher join  to David Cournapeau's project.~\cite{sklearn-history}\\
From this python package are used multiple tools for this work like the different classifiers or the model selection functions.

\subsection{tsfresh}
The motivation of this work is to improve the later obtained classification results for the NTL classification with the aid of a different approach in the feature processing. Consensus of the team member was to process the data set differently as usual with the library tsfresh. \\
tsfresh is delivered as a Python package and is published under MIT licence \cite{ts-licence}. It's developed by Maximilian Christ and Michael Feindt from  Blue Yonder GmbH and  Andreas W. Kempa-Liehr from the  University of Auckland~\cite[1]{fresh-paper}.\\
The main task of the library tsfresh is the calculation of a huge number of  features extracted from time series. Moreover, the library is able to evaluate  the calculated features and select the most important of them. It's possible to start working with this library with a small knowledge about the data and the objectives to be achieved.~\cite[1]{fresh-paper}\\

Principally there are two sub-packages which are the most important ones, the first package is called \textit{feature\_extraction} which purpose it is to calculate the characteristics of feeded data. The other package perform a probability  calculation for every feature, after this the Benjamini Hochberg procedure is used for the selection of the most important features.~\cite{github-feature}\\
It has been developed for large data sets which often occurs in the industry 4.0, in IoT (Internet of Things) or accordingly this work, in machine learning. In such cases a useful ability is the scalability of the FRESH (FeatuRe Extraction based on Scalable Hypothesis
tests) algorithm for ample data sets.~\cite[2-3]{fresh-paper}

\subsubsection{Feature extraction}
tsfresh brings an extensive set of algorithms especially for the feature extraction of time series, afterwards the most ample features are selected.~\cite[3]{fresh-paper}
These computed features represents a reduced  dimensionality of time series which allotted for the training of diverse classifiers. A different approach of time series classification were different types of shape-based methods for the training.~\cite[4-5]{fresh-paper}\\

For the feature calculating  in tsfresh the developers used around 55 algorithms from the collection of Fulcher and  Jones~\cite{fulcher-algo} and Nun et al.~\cite{nun-algo}. \newline
The selection of different algorithms for the feature calculation comprises algorithms for:
\begin{itemize}
	\item Summary statistics, such as maximum, variance or kurtosis.
	\item Characteristics from sample distribution, such as absolute energy, whether a distribution is symmetric or the number of data points above the median.
	\item Observed dynamics, such as fast Fourier transformation coefficients, autocorrelation lags or mean value of the second derivative.
\end{itemize}
The complete list of standard function is available in the tsfresh module description~\cite{readdoc-featex}.

\subsubsection{Feature selection}
The second main step of the feature processing in tsfresh is the feature filtering. It is important that features have a good ration of robustness, which is helpful against outliers and other effects as well, and meaningfulness.\\
To identify meaningful features the library use several hypothesis tests to calculate a  p-value for every feature. The advantage of these hypothesis tests are the robustness of the whole procedure. In contrast another method would be the principal component analysis (PCA). However real world data sets are noisy, which can lead to poor performance of PCA~\cite{pca-fu}. Therefore in this work PCA was not used. \\
The hypothesis tests are applied on every previously extracted feature, the resulting p-values show how much a specific feature is relevant for the final prediction of the target label $y$.  The usage of multiple hypothesis tests offers a better result. There are  different hypothesis tests available in the library, Fisher's exact test, Kolmogorov-Smirnov test for binary and non-binary features and the Kendal rank test.~\cite[5-8]{fresh-paper}\\

Subsequently the following task is to choose which features to adopt and which to reject. This happens with aid of the Benjamini Hochberg procedure,  this algorithm selects based on the p-values and depending on the false discovery rate which features are relevant for the prediction.

\chapter{Data}

\section{Overview of the data set}
\label{subsec:overview_data}
For this work, the data sets are provided by CHOICE Technologies Holding S\`{a}rl, which creates solutions for NTL prediction for their customers. The collaboration between the University of Luxembourg  and this company has the goal to improve their prediction quality for their customers. One of these customers is a Brazilian energy company. The data come from this power company and are used for this work.\\
These data encompass the monthly electricity meter reading from the whole region, about 197 million entries, which contain the user ID, date of recording, an entry for the recorded power consumption, an entry for the charged power consumption and many other records. This data set provides all the power consumption entries for this work. The consumption is measured in kWh.\\
A second data set about customer inspections contains about 800 thousands entries with different values like the date of  inspection, user ID, the result of the inspection and many others. 

\section{Coherence between tsfresh and the data set}
Applying the tsfresh library on the raw data set is impossible, due to a different structure of the data set towards the needs of the library. Also, a preprocessing is required to provide coveted data for the later following feature extraction.\\
Therefore, this work is to preprocess the data set, to apply tsfresh on the selected data, to transform the results of tsfresh to a convenient a data format, and finally, to evaluate the received results. For the last topic, different settings were used to obtain various results and to evaluate those.

\subsection{Loading: constraints of selecting data, parallelization, etc.}
For the preprocessing, processing itself, and the evaluation, diverse Python modules exist. One of the  most important modules for preprocessing is \textit{modi\_data}: At the end, as a result, the module returns all consumptions per customer in a row for a specific time range. Additionally, the latest consumptions of the time series are matched with the latest inspection of the corresponding customer. Only the last inspection of only the customers that have ever been inspected will be used for this preprocessing. The entries in the inspection data set for the result of realized inspection will later represent the target vector $y$, which indicates whether NTLs have happened or not.\\
The reason for using the latest inspection for every customer is to use the most up-to-date data and the longest possible time range.\\
At this stage, there is a time series for the latest inspection of every  inspected customer. \\
The subsequent step is to ensure that all time series have the submitted length and are sustained. This particularly means that no skip in the time series is permitted; the consecutive order is unalterable. Incongruous time series will be removed; this means of course that the appertaining target vector \textit{y} as well as the important customer IDs and the dates of consumption series will be modified.  The compiled dates are important for later sorting purposes. \\
The module \textit{modi\_data} needs the assistance of the module \textit{processData}, which is able to load and save data.
\\
The next step is performed with \textit{tsfreshwrapper}, which transforms the previous results to a tsfresh-specific format and starts the execution.\\
With \textit{slice\_extracted\_features}, it is possible to make different combinations of the calculated features for later evaluation purposes.\\
At the end, different classifiers are trained with the help of \textit{main\_cv} and will point out the quality of the features.
\\
This was the simplified basic logic function of the diverse modules. In the following, these will be explained in detail.

\subsubsection{Modi\_data}
The realisation of the previously made plan for the preprocessing of the data requires some additional explanation. In the following, important steps in the script are clarified.

The module \textit{modi\_data} is responsible for performing a major part of the data set preprocessing. The module contains only one function, which accepts as parameters the length of the prospective time series and a boolean, which indicates to save the results; this is useful for a subsequent processing of the results.  Another parameter is the \textit{consumptions\_column}, with which the different recording types, see Section~\ref{subsec:overview_data}, can be selected. The last parameter of the signature is \textit{drop\_all\_zero\_consumptions}; this boolean is responsible for removing time series whose entries are all zeros from the output.

\begin{figure}[h!]
	\center
	\includegraphics[scale=0.2]{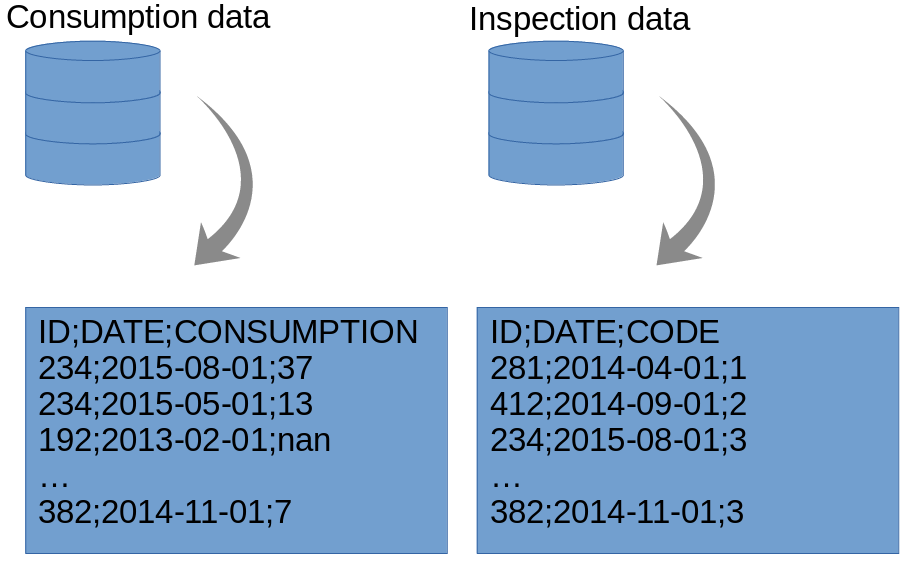}
	\caption{Read data}
	\label{fig:readin}
\end{figure}

At first, see Figure~\ref{fig:readin}, the data, like the user ID, the date of the record, the chosen consumption type and the inspection result, are read in with the help of the module \textit{proxessData}.

\begin{figure}[h!]
	\center
	\includegraphics[scale=0.3]{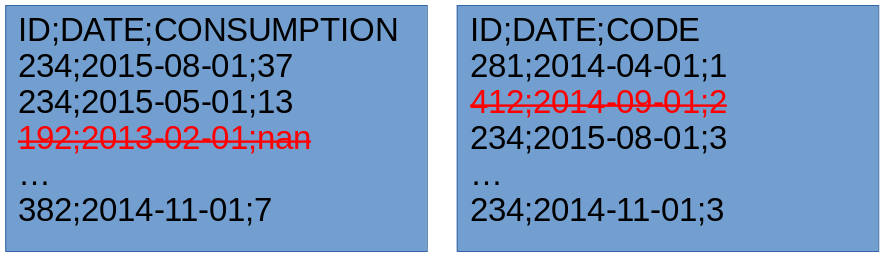}
	\caption{Convert and drop data}
	\label{fig:drop}
\end{figure}

The next step is to drop the rows with unusable values and the rows with wrong entries in the inspection result. Afterwards, the read data is converted to the correct data type. Therefore see Figure~\ref{fig:drop}.

\begin{figure}[h!]
	\center
	\includegraphics[scale=0.25]{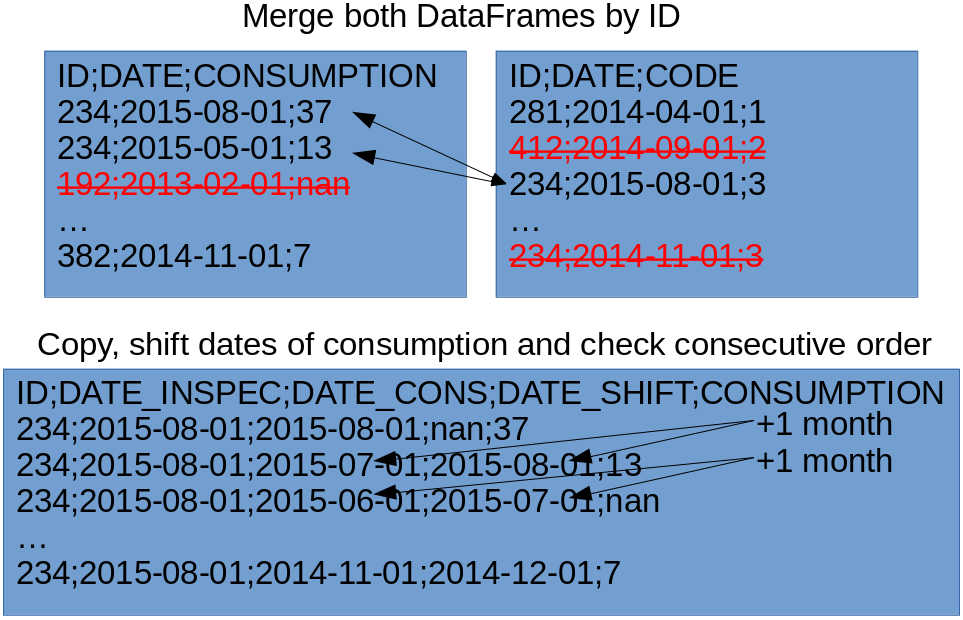}
	\caption{Merge, shift and check sequence}
	\label{fig:merge}
\end{figure}

Following this, the latest inspection is determined and merged as an inner-join with the consumptions. As a result of this, many time series are of insufficient length or have gaps in their rows. To check for consecutive time series, the column with the dates is copied and shifted by one. This makes it easy to compare the current date with the next one. Depicted by Figure~\ref{fig:merge}.

Next, a pandas \texttt{groupBy} object is created, which contains every time series as a group. One interesting step is to check these time series, as this happens parallelized with a Python list. The chunk size is calculated from the number of groups divided by amount of processor threads minus one. It is so calculated because one list element must contain the remainder.\\
By extracting the groups from the \texttt{groupBy} object, concatenating them to a new smaller \texttt{groupBy} object and appending this as a new element to the list, the following code is executed:
\\
\renewcommand{\lstlistingname}{Source Code}
\renewcommand{\lstlistlistingname}{List of \lstlistingname s}
\begin{lstlisting}[label=filllist, caption=Filling the list for parallelized computing, language=myPython, firstline=3, lastline=6]
data_parts = []
    for jumz in range(0, len(gr_ob), chunk_size):
        data_parts.append( pd.concat( [ gr_ob.get_group(group) for i,group in enumerate( gr_ob.groups) if (i>= jumz) and (i < jumz+chunk_size)  ] ).groupby('ID_UC') )
    print len(data_parts)
\end{lstlisting}
\ \\
Up next is an enumerable list available for the parallelized execution. This way is chosen for the parallelization because it is hard to use other parallelization methods. For example, the \texttt{Pool} function requires a pickable data object. The Python \textit{pickle} module helps to transform an object to a byte stream~\cite{python-pickle}.  This byte stream is easy to share between different platforms, to save  and its properties are important for many parallelization methods. However, the method used in this work requires  the data to be split into individual groups because pandas objects are not pickable.\\
The actual parallelization is performed with pipes: every pipe applies the \texttt{inner\_group} function on the time series from the list elements. This internal function  checks the correct length and  consecutive numbering of the dates of the time series.\\
The retained time series, and the other values like the userID, are inserted in a shared queue object, which provides the possibility for multiple access~\cite{queue-manager}. After the processing, the time series and the other entries are pulled out of the queue and put together to new NumPy \texttt{arrays}. This procedure will not lead to any problems because every group element is an atomic unit. A different order of the single time series does not make any difference to the overall outcome.\\
The last step is to return and optionally save the results.\\

A first revision of the code with the objective to increase execution performance was realised with the help of the Dask collection. Thereby, many data types and functions were replaced in this code. Also the code itself was modified to fit to the new package. These modifications result in partially five percentage points higher execution speed than before.

\subsubsection{ProcessData}
One very helpful file is \textit{processData}, which is responsible for loading different data files and saving the results.
First, it is important to mention that the loading functions of the different data files always expect the files with same name and location.
However, a different user name of the file system represents no problem.
\\

A function that is used in this work is \texttt{get\_raw\_consumptions}. This Python function reads from the files, with the meter readings, the two columns for the user ID and the date of the recording. Also it is possible to pass additional column labels with the parameter \texttt{extra\_columns}, for example to discriminate between the measured and the charged electrical power consumption. During the feed-in of the data, the columns with the consumption dates are converted into a pandas date format. The return value of this read-in function is a pandas \texttt{DataFrame}.

In the same way, \texttt{get\_raw\_inspections} works, whose task is to read the entries from the file with the inspections of the customers. Besides the user ID and the date of inspection, it also reads the inspection result. The return value is a pandas \texttt{DataFrame}, too.

This Python module also owns functions to save the computed results. The function \texttt{save\_assessed\_dataframe} has the job of saving resulting \texttt{DataFrame}s as a simple csv file, where every entry is separated by a semicolon. Besides the \texttt{DataFrame} itself, it accepts an optional string that is added to the file name to describe the file. In addition, a unique time stamp is concatenated to later discriminate the different files.

During the data preprocessing, NumPy arrays emerge that contain different results like the consumptions, the target vector, the IDs and the dates of the consumptions. These have to be saved, and therefore \texttt{save\_result\_np} exists. This function takes, of course, the array and a parameter called \texttt{fmt}, which represents the data format, such as integer or floats with a different length. The last parameter is responsible for the name of the file; also, a time stamp is added. The saved file corresponds the csv file type.

\subsubsection{Tsfreshwrapper}
After the preprocessing with \textit{modi\_data},  \textit{tsfreshwrapper} follows, which is a wrapper function for the actual tsfresh library. This function transforms the standard format of the data, one time series per line, to a  tsfresh-acceptable format. Afterwards, applying tsfresh on the data, \textit{tsfreshwrapper} is able to save the results before and after the feature filtering, which is useful for later purposes. 

\subsubsection{Slice\_extracted\_features}
In this work, diverse types of features calculation algorithms are used. 
Some tsfresh itself provides, the  in this work called standard features or generic  time series features (GTS). The daily average (AVG) and the features that are especially created for the NTL detection, and the difference features (DIF), for further information see Section~\ref{sec:dif}.
To compare  the gained features from each of these algorithms, it is necessary to split them. Since all of these different feature types are computed at once with tsfresh, and the result is a \texttt{DataFrame} with all the features, is it important to have the ability to separate the different feature types for the evaluations following afterwards.\\
For this work, the module \textit{slice\_extracted\_features} is created. The signature of the primary function \texttt{slice\_features} accepts, besides the file name, the length of the time series and the number of processes, a variable amount of enums. For every feature type an enum exists, such as \texttt{DIFFERENCE\_FEATURES}, \texttt{DAILY\_AVERAGE} and \texttt{PURE\_TSFRESH}. With these enums it is possible to name every arbitrary combination of features and compute these. Furthermore, it is feasible to perform a feature selection after calculating the combinations.  For every combination a new \texttt{DataFrame} will be saved.\\

At the current state, the module is only able to handle these three feature types. An improvement could be the read in of the different feature calculation algorithms from the \textit{feature\_calculators} file, which contains all the algorithms in tsfresh. In addition, it would be useful to use different Python decorators for the different feature types. That would have the advantage of being able to use an endless number of different feature types.

\subsubsection{Main\_cv}
\label{subsec:maincv}
To evaluate the different types of features with diverse classifiers, it is important to train them first and subsequently to test them. However, training the machine learning model itself is not enough; it is also important to determine the hyperparameter of every model. Therefore, the function \texttt{RandomizedSearchCV} from \texttt{scikit-learn} is used. For a more detailed description of the model training, see Section~\ref{sec:methods}.
\\
The function \texttt{determine\_classifier\_parameter} receives a file with the previously calculated features and a file with the target label. After converting the data into a NumPy \texttt{matrix}, the \texttt{preprocessing.scale()} command standardizes the shape of the distribution~\cite{sklearn-preprocessing}. \\
With these data, the function \texttt{run\_randomsearch} performs a training for every classifier. This function determines the best hyperparameters for the classifiers and the mean ROC-AUC (see Section~\ref{sec:auc}) and its inherent standard deviation.\\
This outcome results after 1k iterations, which are composed of the number of iterations for \texttt{RandomizedSearchCV} and the size of the k-fold cross-validation, which triggers the corresponding number of iterations for the fold size. The whole training and parameter ranking lasts over four days. The training of the gradientboosted classifier takes the longest time. For a future productive application of this module, it is only required to train the best performing classifier; this will reduce the whole training time.

\chapter{Detection of non-technical losses}

\section{Additional feature calculation functions}
\label{sec:dif}
The tsfresh library is designed for the extraction and filtering of features from different kinds of data or even time series. Therefore, developers promote it as a useful tool for every case where a lack of domain knowledge exists.\\
The cooperation between the SEDAN research group and the partner company already produced a certain knowledge in the field of NTL detection and hence they have published different papers. A helpful paper is about information extraction from provider-independent data~\cite{distilling-jorge}. The feature calculation functions presented in this paper are only developed for the time series classification problem in the domain of NTL detection. This fits together excellently with this work for the reason that the tsfresh library provides the possibility to implement user-specific feature calculation algorithms. \\
This offers the opportunity to implement the extra features from Meira et. al. in the designated file \textit{feature\_calculators} of the library.\\
For the following algorithms it is assumed that the consumption time series is consecutive and has a length of $N$ months and is described as:
\begin{align}
C^{(m)} = [C^{(m)}_0, ..., C^{(m)}_{N-1}],
\end{align}
where $C^{(m)}_{N-1}$ is the latest meter reading before the inspection.\\

The first of three difference functions is called fixed interval. Together with $K \in \{3, 6, 12\}$, it computes  the difference between the current consumption and the mean consumption in a period directly before a meter reading. After applying the function, it generates $3\times (N - 12)$ features for the data set. 
\begin{align}
\operatorname{fixed\_interval}_d^{(m)} = C_d^{(m)} - \frac{1}{K} \times \sum^{d-1}_{k=d-K} C_{k}^{(m)},
\end{align}

Another algorithm for the calculation of features is the intra-year difference
\begin{align}
\operatorname{intra\_year}_d^{(m)} = C_d^{(m)} - C_{d-K}^{(m)},
\end{align}
with $K = 12$, which is the difference of consumption to the consumption in the same month of the previous year. Overall, the function returns $N - 12$ different features.

The intra-year seasonal difference
\begin{align}
\operatorname{intra\_year\_seasonal}_d^{(m)} = C_d^{(m)} - \frac{1}{3} \times\sum^{d-K+1}_{k=d-K-1} C_{k}^{(m)},
\end{align}
for $K = 12$, is the change of consumption of the mean of three months in the previous year to the current consumption. Altogether, the intra-year seasonal difference delivers $N - 13$  features. 
\newline
The previously used daily average feature, which also serve as baseline feature for purposes of comparison, is for the month $d$ for a customer $m$ in kWh:
\begin{align}
\operatorname{daily\_avg}_d^{(m)} = \frac{C_d^{(m)}}{R^{(m)}_{d} - R^{(m)}_{d-1}}.
\end{align}
$C_d^{(m)}$ is the consumption between the meter readings $R^{(m)}_d$ and $R^{(m)}_{d-1}$, where $d$ is the current month and $d-1$ the month before. The difference between $R^{(m)}_d$ and $R^{(m)}_{d-1}$ results in the days between the two meter readings. Concerning this calculation, the function returns only $N-1$ features. This feature is  common in the field of NTL detection~\cite{nagi2008detection, nagi2010nontechnical, nagi2011improving}.

\section{Modification of tsfresh}
The already previously used daily average function as well as the three new functions were implemented in the \textit{feature\_\-calculators}.
tsfresh provides three different use cases for the implementation; these are represented by three different decorators. A Python decorator is related to annotations in Java, which also start with the @ symbol. In general, their task is to modify elements of the language, for example, functions or whole classes~\cite{python-anno}. In the case of tsfresh, the decorators distinguish between the three deployment types. 

To calculate a single feature without any parameter, the \textit{aggregate feature without parameters} is suggested.
The second function type, \textit{aggregate features with parameter}, uses the provided parameter from the \textit{settings} file.
The last one and the type used for this work is called \textit{apply} and it is able to calculate multiple features at the same time  for different  parameters.

It is also possible to utilize parameters for the calculations. The parameters are provided in  the \textit{settings} file. Inside the constructor of the function \texttt{Feature\-Extraction\-Settings} there is a dictionary, \texttt{name\_to\_param.update}, which contains the parameters for  the functions that actually require parameters~\cite{readdoc-custom}. 
The implementations of the custom feature calculators are provided in \textit{feature\_calculators}. \\

However, to code the daily average function, a bit more effort was needed.
As already mentioned in the explanation of the daily average function, the consumption $C_d^{(m)}$ is divided by the number of days between both meter readings; to compute these days, the specific dates for every consumption are required. Albeit tsfresh uses the optional parameter \texttt{column\_sort}  only to sort the data, it drops the column with the dates.\\
This leads to the problem that tsfresh does not conduct the column with the consumption dates to the daily average function. As a result, no dates are available to calculate the divisor. A chance would be to use the business month with 30 days as a divisor, but in order to avoid inaccuracies caused by the different lengths of the months or due to leap years, a more precise way, nevertheless, is to use the original dates. To reach this objective, it is necessary to modify tsfresh itself. 

The important function flow regarding the columns with the dates is the following: first, passing the parameters to the main extraction function  \texttt{extract\_features}  in \textit{extraction.py}, a normalization function \texttt{normalize\_\-input\_\-to\_\-inter\-nal\_\-re\-presen\-tation} follows, which is able to change the input data into a uniform structure and remove all unnecessary columns except  the columns for value and ID.
The column with the dates is only used to sort the data and then is dropped, see Section~\ref{lst:dropformer}.
\\
\renewcommand{\lstlistingname}{Source Code}
\renewcommand{\lstlistlistingname}{List of \lstlistingname s}
\begin{lstlisting}[label=lst:dropformer, caption=Former implementation, language=myPython, firstline=1, lastline=1]
kind_to_df_map[kind] = kind_to_df_map[kind].sort_values(column_sort).drop(column_sort, axis=1)
\end{lstlisting}
\ \\
Subsequently, the internal function \texttt{\_extract\_\-features\_\-for\_\-one\_\-time\_\-series}, which is responsible to extract the features from the data frame, calls the function \texttt{get\_apply\_functions}.
This latter function creates a list with all feature calculation algorithms, the column prefix and the parameter belonging to the decorator \textit{apply}.
This list is returned to its previous function, and every extraction function is applied to the data. After some further steps, tsfresh presents the results of the feature extraction.\\

To ensure that the dates are passed through to the daily average function, some modifications are needed in tsfresh.\\
In order to avoid dropping the dates through the normalization function, the suffix \texttt{.drop(column\_sort, axis=1)} has to be removed.
Then the \texttt{get\_apply\_functions} must be modified, which is called by the internal feature extraction function \texttt{\_extract\_\-features\_\-for\_\-one\_\-time\_\-series}.
This function gathers the different feature calculation algorithms from the \textit{feature\_calculator} file, the column prefix and the parameter from \textit{settings}, and gets as another value the name of the feature calculation algorithm.
This returns the lists to their top function, the internal feature calculator, the \texttt{\_extract\_\-features\_\-for\_\-one\_\-time\_\-series}. Next, a selection is made there, between the normal functions and the daily average function.
The daily average function additionally gets the previously retained \texttt{column\_sort} data and is applied to these data. Thus, the daily average function can use the dates to calculate the days for the divisor.\\
All these modifications that were made on tsfresh are available as a fork from the original tsfresh project under: \url{https://github.com/hargorde/tsfresh}.
For later modifications, a more general approach would be to use a new decorator for all functions that require the dates of the consumptions. 

\chapter{Evaluation}
\section{Methods}
\label{sec:methods}
For the evaluation of the different features, the responsible module \texttt{main\_cv}, see Section~\ref{subsec:maincv}, uses \texttt{RandomizedSearchCV} with a cross-validation (cv) of the type k-fold. Both are described in detail in the following. \\

\subsection{Cross-validation}
Basically, to train a machine learning model, the provided data set is split into a training data set and a test data set. The trainings set is used for training and the test set verifies the model afterwards. In order to avoid overfitting to a specific set, it is important to use different sets.\\
For the determination of the hyperparameters, an extra set is required: the validation set. Within the library \texttt{scikit-learn}, the {\hp}s must be set before the training with help of the constructor. This means the training is still performed with the training set; then follows the validation set to check that the {\hp}s fit and when the {\hp}s fit; then follows the verification with the test data set.\\
However, splitting the data into three different sets reduces the amount of available data. This can partially influence the machine learning model; to avoid this effect, cross-validation (cv) is predestined for such issues.

A common type of cross-validation is the k-fold, which uses a collective data set for training and validation; only  for the test an extra set is required.  In \texttt{scikit-learn}, the k-fold  method divides the training set into $k$ smaller sets, and the size of $k$ is set by the parameter \texttt{n\_splits}. 
From these subsets $k-1$ are used for the training of the machine learning model. The remaining fold is used for the validation. This happens $k$ times: for each combination of the subsets, a training and a validation is performed. \\
Although the k-fold cross-validation needs some computational power, it is useful for small sets.
Besides the k-fold  cross-validators, others exist, such as Leave One Out, Leave P Out.
A cross-validation function  provided by the library is  \texttt{cross\_val\_score}. This function  takes the model,  the data set, target set, the number of folds, and optionally a scoring  function to evaluate the training results.  In this work, the scoring method used is AUC.~\cite{sklearn-cv} 

\subsection{Randomized parameter optimization}
The \textit{randomized parameter optimization} optimizes of the hyperparamters. Thereby, a dictionary with a set of different hyperparameters is defined for this procedure and passed to the function \texttt{RandomizedSearchCV}. The function creates a set with randomly picked hyperparameters. The number of different combination is determined with the parameter \texttt{n\_iter}. If the \texttt{n\_iter} parameter is increased, the quality of the hyperparmeter search also increases, but the calculation period rises as well. Optionally it is possible to enter a cross-validation method like k-folds.
At the end, the best-fitting hyperparameters are retained.~\cite{sklearn-grid}

\section{AUC}
\label{sec:auc}
The data set used for this work is, as already mentioned, a real world data set; this means, the data contains noise and outliers. Also, a special challenge is the handling of the irregularities itself, because they are imbalanced~\cite[8]{survey-glauner}. This becomes clear after a first processing of the data for the measured consumption: There are 150,700 entries that refer to non-NTL, and 50,229 entries that are NTL. The outcome confirms that  both classes NTL and non-NTL are imbalanced, which is clearly shown by a NTL rate of 33.3\% regarding the whole data.\\

Normally, to rate the results of a prediction, a performance measure such as accuracy or precession is often used.  At first, measurement methods like accuracy and precision might explain things, but they have problems with imbalanced data sets. 
\\
The accuracy is the true positives (TP) plus the true negatives (TN) over the number of the whole set: $ACC = \frac{TP+TN}{TP+TN+FP+FN}$. The precision is like this: $P = \frac{TP}{TP+FP}$. Another function is the recall, which is $R = \frac{TP}{TP+FN}$ ~\cite{survey-glauner, glauner2016large, sklearn-precision}.\\

The following example from a survey about NTL detection~\cite[8]{survey-glauner} shows very clearly the disadvantage from these scores. There is an example data set with 100 customers of which 99 are non-NTL. If a classifier predicts non-NTL for all the entries, the resulting accuracy would be 99\%.\\
In contrast, would the classifier always predict NTL, the recall would be 100\%.\\ 
In the first case, the classifier will never detect NTL despite a high accuracy.
The second case will find the NTL, but forces too many inspections to be carried out by inspectors. This behaviour leads to a rise in costs.
\\
These results show that it is important to choose another evaluation function for the trained classifiers, a function that considers the imbalance of both classes \cite{survey-glauner}.\\

A better evaluation for the output of the different classifiers is to plot the recall and the opposite of the recall against each other. This results in a receiver operating
characteristic (ROC) curve. The area under the curve (AUC), i.e., the ROC curve, is scored from 0 to 1, where 1 means every prediction is correct and 0 points out that no prediction results are correct. Every score over 0.5 is better than random guessing. In this work, the function used for the AUC is: $AUC = \frac{Recall+Specificity}{2}$ ~\cite{glauner2016large}.  

\section{Results}
At the end, the program was applied to the data set for the measured and billed consumption, at first transforming and collating the data, then matching the entries and checking the consumption time series.\\
Subsequently, the tsfresh library, with its extra features, was applied on the primed data and extracted many different features.\\
After the calculation of the features, the different feature types were separated with the \texttt{slice\_features} function and diverse combination of these were created.\\
A first machine learning model training with scoring started  with these unfiltered features. \\
The features were then filtered with the help of tsfresh regarding their importance for the target label $y$. After this, a model training with an evaluation of their prediction results starts.\\

Table~\ref{table:features} shows the number of gained different feature types like the daily average (AVG), generic time series (GTS), the fixed interval, intra-year difference and intra-year seasonal difference, which are also called difference features (DIF).
\begin{table}[h]
	\renewcommand{\arraystretch}{1.3}
	\caption{Number of features before and after selection.}
	\label{table:features}
	\centering
	\begin{tabular}{|l|c|c|c|}
		\hline
		\multirow{ 2}{*}{Name} & \multirow{ 2}{*}{\#Features} & \multicolumn{2}{|c|}{\#Retained Features} \\
		\cline{3-4}
		& & Measured & Billed \\
		\hline
		\hline
		Daily average (AVG) & 23 & 18 & 23 \\
		\hline
		Fixed interval & 36 & 34 & 36 \\
		\hline
		Generic time series (GTS) & 222 & 162 & 201 \\
		\hline
		Intra-year difference & 12 & 12 & 12 \\
		\hline
		Intra-year seasonal difference & 11 & 11 & 11 \\
		\hline
		\hline
		Total & 304 & 237 & 283 \\
		\hline
	\end{tabular}
\end{table}
The column \textit{\#Features} stands for the number of features before the feature selection. The  other column, \textit{\#Retainded Features}, with its sub-labels for the measured and actually billed consumption, contains the number of different features after a feature filtering.\\
After the features extraction, tsfresh returns 304 features for both consumption verdicts, because the library applies all feature calculation algorithms with all parameters on the data. The following filtering retains 237 features for the measured consumption and 283 features for the billed consumption. 
It turns out that the features that were only made for the NTL detection are the best obtained ones. Only two fixed interval features, with a window of $K = 3$, were dropped from the measured consumption; probably it coheres due the short time range. For NTL detection, the common daily average features are completely retained for the billed data, but from the features for the measured data, only 18 out of 23 are kept. An interesting fact is that the five dropped features represent the six oldest months of a time series with a total length of 24 months. It seems that for the feature selection only the latest 18 months are important.\\
Both consumption types show in comparison that for the billed data many more features  are retained; a reason could be the major scope of entries.
During the preprocessing of the measured data, around 135 thousand consumption time series are dropped, because they were a complete zero sequence. The different size and the different data also have an impact on the generic time series algorithms, where 73\% and 91\%  of the features are retained for the measured and billed data. Another reason is that the generic time series features are not made for NTL detection and thereby produce features of lower quality.\\

The next Table~\ref{table:resmedido} is created for the measured consumption. It compares the different classifiers and the feature type combinations.
\begin{table*}[!ht]
	\renewcommand{\arraystretch}{1.3}
	\setlength{\tabcolsep}{0pt}
	\caption{Test Performance of Classifiers on Features from Measured Consumption Data.}
	\begin{minipage} {0.95\textwidth}
		\label{table:resmedido}
		\begin{flushleft}
			\begin{tabular}{|l|c|c|c|c|c|c|c|c|}
				\hline
				\multirow{ 2}{*}{Clf.} & \multicolumn{2}{|c|}{GTS} & \multicolumn{2}{|c|}{AVG} & \multicolumn{2}{|c|}{DIF} & \multicolumn{2}{|c|}{GTS+AVG} \\
				\cline{2-9}
				& $X_{all}$ & $X_{ret}$ & $X_{all}$ & $X_{ret}$ & $X_{all}$ & $X_{ret}$ & $X_{all}$ & $X_{ret}$ \\
				\hline
				\hline
				DT & 0.64544 & \cellcolor[gray]{0.9} 0.64625 & \cellcolor[gray]{0.9} 0.64037 & 0.63985 & 0.63730 & \cellcolor[gray]{0.9} 0.63792 & \cellcolor[gray]{0.9} 0.64712 & 0.64705 \\
				\hline
				RF & $0.65665^{c}$ & \cellcolor[gray]{0.9} $0.65726^{c}$ & $0.65083^{c}$ & \cellcolor[gray]{0.9} $0.65248^{c}$ & \cellcolor[gray]{0.9} $0.65529^{c}$ & $0.65459^{c}$ & $0.65800^{c}$ & \cellcolor[gray]{0.9} $0.65835^{c}$ \\
				\hline
				GBT & \cellcolor[gray]{0.9} 0.63149 & 0.63125 & \cellcolor[gray]{0.9}  0.63234 & 0.63186 & 0.62869 & \cellcolor[gray]{0.9}  0.63019 & 0.63262 & \cellcolor[gray]{0.9} 0.63322 \\
				\hline
				LSVM & \cellcolor[gray]{0.9} 0.63696 & 0.63656 & \cellcolor[gray]{0.9} 0.54982 & 0.54933 & 0.55749 & \cellcolor[gray]{0.9} 0.55843 & 0.63725 & \cellcolor[gray]{0.9} 0.63689\\
				\hline
			\end{tabular}
		
		\end{flushleft}
	\end{minipage}
	\\[0.5cm]
	\renewcommand{\arraystretch}{1.3}
	\setlength{\tabcolsep}{0pt}
	\begin{minipage} {0.95\textwidth}
		\begin{flushleft}
			\begin{tabular}{|l|c|c|c|c|c|c|}
				\hline
				\multirow{ 2}{*}{Clf.} &  \multicolumn{2}{|c|}{GTS+DIF} & \multicolumn{2}{|c|}{AVG+DIF} & \multicolumn{2}{|c|}{GTS+AVG+DIF} \\
				\cline{2-7}
				& $X_{all}$ & $X_{ret}$ & $X_{all}$ & $X_{ret}$ & $X_{all}$ & $X_{ret}$  \\
				\hline
				\hline
				DT & 0.64638 & \cellcolor[gray]{0.9} 0.64647 & \cellcolor[gray]{0.9} 0.64348 & 0.64312 & 0.64646 & \cellcolor[gray]{0.9} $0.64765^{f}$\\
				\hline
				RF & \cellcolor[gray]{0.9} $0.65911^{c}$ & $0.65896^{c}$ & \cellcolor[gray]{0.9} $0.65858^{c}$ & $0.65755^{c}$ & $0.65747^{c}$ & \cellcolor[gray]{0.9} $\textbf{0.65977}^{cf}$ \\
				\hline
				GBT & 0.63319 & \cellcolor[gray]{0.9} $0.63358^{f}$ & \cellcolor[gray]{0.9} 0.63261 & 0.63245 & 0.63354 & \cellcolor[gray]{0.9} 0.63355 \\
				\hline
				LSVM & \cellcolor[gray]{0.9} 0.63731 & 0.63693 & 0.57173 & \cellcolor[gray]{0.9} 0.57432 & 0.63728 & \cellcolor[gray]{0.9} $0.63760^{f}$ \\
				\hline
			\end{tabular}
		\end{flushleft}
		Test AUC for combinations of decision tree (DT), random forest (RF), gradient boosted tree (GBT) and linear support vector machine (LSVM) classifiers trained on sets composed of general time series (GTS), daily average (AVG) and difference (DIF) features.\\
		The best overall combination of classifier and feature set is \textbf{highlighted}.\\
		Per combination of classifier and feature set, the better result on either a full feature set ($X_{all})$ or retained feature set ($X_{ret})$ is \colorbox{light-gray}{highlighted}. \\
		$^c$ denotes the best classifier per feature set.\\
		$^f$ denotes the best feature set per classifier.
	\end{minipage}
\end{table*}
To evaluate the work, different classifiers are trained with diverse features like the AVG, GTS and DIF features and combinations of them. The best prediction results are achieved by the random forest for all the different feature combinations. The best features set consists of a selection of all three feature types and gain an AUC of 0.65977. A comparison of the AUC between the unfiltered and the filtered features results in 27 significant differences. In 15 of 27 cases the filtered features perform superior to the unfiltered features.\\

For the billed consumption, see Table~\ref{table:resX2}, the random forest performs most precisely of all machine learning models as well. The best training result is reached after a filtering  of the daily average, generic time series and difference features with an AUC of 0.67356. 
\begin{table*}[!ht]
	\renewcommand{\arraystretch}{1.3}
	\setlength{\tabcolsep}{0pt}
	\caption{Test Performance of Classifiers on Features from Billed Consumption Data.}
	\begin{minipage} {0.85\textwidth}
		\label{table:resX2}
		\begin{flushleft}
			\begin{tabular}{|l|c|c|c|c|c|c|c|c|}
				\hline
				\multirow{ 2}{*}{Clf.} & \multicolumn{2}{|c|}{GTS} & \multicolumn{2}{|c|}{AVG} & \multicolumn{2}{|c|}{DIF} & \multicolumn{2}{|c|}{GTS+AVG}\\
				\cline{2-9}
				& $X_{all}$ & $X_{ret}$ & $X_{all}$ & $X_{ret}$ & $X_{all}$ & $X_{ret}$ & $X_{all}$ & $X_{ret}$  \\
				\hline
				\hline
				DT & 0.65901 & \cellcolor[gray]{0.9} 0.65936 & 0.65626 & \cellcolor[gray]{0.9} 0.65654 & \cellcolor[gray]{0.9} 0.64220 & 0.64169 & 0.66040 & \cellcolor[gray]{0.9} 0.66088  \\
				\hline
				RF & $0.66626^{c}$ & \cellcolor[gray]{0.9} $0.66639^{c}$ & \cellcolor[gray]{0.9} $0.66877^{c}$ & $0.66725^{c}$ & $0.65435^{c}$ & \cellcolor[gray]{0.9} $0.65447^{c}$ & $0.66765^{c}$ & \cellcolor[gray]{0.9} $0.66990^{c}$  \\
				\hline
				GBT & \cellcolor[gray]{0.9} 0.65487 & 0.65479 & 0.65526 & \cellcolor[gray]{0.9} 0.65594 & 0.64044 & \cellcolor[gray]{0.9} 0.64160 & 0.66016 & \cellcolor[gray]{0.9} 0.66021 \\
				\hline
				LSVM & 0.64481 & \cellcolor[gray]{0.9} 0.64484 & \cellcolor[gray]{0.9} 0.60558 & 0.60512 & 0.60512 & 0.60512 & 0.64530 & \cellcolor[gray]{0.9} 0.64533  \\
				\hline
			\end{tabular}
		\end{flushleft}
	\end{minipage}
	\\[0.5cm]
	\renewcommand{\arraystretch}{1.3}
	\setlength{\tabcolsep}{0pt}
	\begin{minipage} {0.85\textwidth}
		\begin{flushleft}
			\begin{tabular}{|l|c|c|c|c|c|c|c|c|c|c|c|c|c|c|}
				\hline
				\multirow{ 2}{*}{Clf.} &  \multicolumn{2}{|c|}{GTS+DIF} & \multicolumn{2}{|c|}{AVG+DIF} & \multicolumn{2}{|c|}{GTS+AVG+DIF} \\
				\cline{2-7}
				& $X_{all}$ & $X_{ret}$ & $X_{all}$ & $X_{ret}$ & $X_{all}$ & $X_{ret}$  \\
				\hline
				\hline
				DT &  0.66213 & \cellcolor[gray]{0.9} 0.66230 & \cellcolor[gray]{0.9} 0.65982 & 0.65953 & 0.66436 & \cellcolor[gray]{0.9} $0.66494^{f}$ \\
				\hline
				RF & $0.66783^{c}$ & \cellcolor[gray]{0.9} $0.67070^{c}$ & $0.67254^{c}$ & \cellcolor[gray]{0.9} $0.67338^{c}$ & $0.67274^{c}$ & \cellcolor[gray]{0.9} $\textbf{0.67356}^{cf}$ \\
				\hline
				GBT & 0.66110 & 0.66110 & \cellcolor[gray]{0.9} 0.66384 & 0.66359 & 0.66503 & \cellcolor[gray]{0.9} $0.66547^{f}$ \\
				\hline
				LSVM &  0.64511 & \cellcolor[gray]{0.9} 0.64520 & 0.61979 & \cellcolor[gray]{0.9} 0.62032 & \cellcolor[gray]{0.9} $0.64565^{f}$ & 0.64558 \\
				\hline
			\end{tabular}
		\end{flushleft}
		Test AUC for combinations of decision tree (DT), random forest (RF), gradient boosted tree (GBT) and linear support vector machine (LSVM) classifiers trained on sets composed of general time series (GTS), daily average (AVG) and difference (DIF) features.\\
		The best overall combination of classifier and feature set is \textbf{highlighted}.\\
		Per combination of classifier and feature set, the better result on either a full feature set ($X_{all})$ or retained feature set ($X_{ret})$ is \colorbox{light-gray}{highlighted}. If a classifier performs the same  on both feature set, none is highlighted. \\
		$^c$ denotes the best classifier per feature set.\\
		$^f$ denotes the best feature set per classifier.
	\end{minipage}
\end{table*}

In the most cases, the filtered features leads to better prediction results. A relevant difference between the all features and the retained features appears in 26 cases. From these 26 cases, 19 times the filtered features delivered a better AUC.\\
\newline
Overall, it is worthwhile to mention that the AUC rises with more feature sets; this applies to both test series. The prediction quality for single feature sets is the lowest; combinations with two sets lead to a slight increase and the best quality prediction has a combination of all three.

\clearpage

\section{Discussion}
This work shows that the newly extracted features from the data lead to a better prediction quality than the former daily average features. The AUC with the combination of all the features gained is just slightly better than the AUC of the daily average features. A quite important cause for this behaviour is that the provided data set is a real world data set, which contains noise, outliers, or even errors in the data through input error.  However, even a small improvement in the NTL detection would cause to a reduction of the cost for carrying out inspections for the customers. Furthermore, the energy company would be able to increase the stability of the electrical power grid by decreasing the NTL cases. Also, the partner company of this project would have the benefit of raising their equity through the improvement of their products. \\
Surprisingly, the gradient boosted tree always delivered poor classification results. Contrarily, in other papers the classifier showed good prediction quality~\cite{roe2005boosted, chen2016xgboost}. This could be explained with the ``no free lunch'' (NFL) theorem, which points out that the best machine learning model does not exist~\cite{wolpert1996lack}.

\chapter{Conclusion}
Regarding the previously defined objectives of this work, the goal is achieved. \\
With the preprocessing, the tsfresh library, and the machine learning model training, it was possible to obtain more from the consumption time series. This project helped to achieve more meaningful features, also with the help of the extra added features (DIF). The model training points out the best combination of the feature types and the best classifier. \\
A comparison between the formerly used daily average features and the now obtained features shows a slight improvement in the quality of the features. However, the reachable quality is limited because real world data were used in this piece of work.
Besides the improvement of the NTL prediction, this work delivered a framework for feature extraction that is able to incorporate further algorithms especially for NTL detection. \\


\listoftables
\listoffigures
\lstlistoflistings
\printbibliography
\end{document}